# OPTIMIZATION OF A REAL-TIME WAVELET-BASED ALGORITHM FOR IMPROVING SPEECH INTELLIGIBILITY

Tianqu Kang, Anh-Dung Dinh, Binghong Wang, Tianyuan Du, Yijia Chen, and Kevin Chau

Hong Kong University of Science and Technology, Hong Kong, China

tkang@connect.ust.hk, eekchau@ust.hk

## ABSTRACT

The optimization of a wavelet-based algorithm to improve speech intelligibility along with the full data set and results are reported. The discrete-time speech signal is split into frequency sub-bands via a multi-level discrete wavelet transform. Various gains are applied to the sub-band signals before they are recombined to form a modified version of the speech. The sub-band gains are adjusted while keeping the overall signal energy unchanged, and the speech intelligibility under various background interference and simulated hearing loss conditions is enhanced and evaluated objectively and quantitatively using Google Speech-to-Text transcription. A universal set of sub-band gains can work over a range of noise-to-signal ratios up to 4.8 dB. For noise-free speech, overall intelligibility is improved, and the Google transcription accuracy is increased by 16.9 percentage points on average and 86.7 maximum by reallocating the spectral energy toward the mid-frequency sub-bands. For speech already corrupted by noise, improving intelligibility is challenging but still realizable with an increased transcription accuracy of 9.5 percentage points on average and 71.4 maximum. The proposed algorithm is implementable for real-time speech processing and comparatively simpler than previous algorithms. Potential applications include speech enhancement, hearing aids, machine listening, and a better understanding of speech intelligibility.

## 1. INTRODUCTION

Speech intelligibility refers to the comprehensibility of speech, and is comprised of the speaker, the listener, and the listening environment. Factors that affect speech intelligibility include the clarity and content of the speech, whether the speaker and listener are native or proficient speakers of the language, and whether the listener is impaired by hearing loss. The listening environment heavily affects intelligibility as well, e.g., excessive reverberations, distortions, background noise, and competing speech (cf. the cocktail party effect [1]) all degrade intelligibility.

Considerable research has been conducted to understand and improve speech intelligibility in a noisy environment and for people with hearing loss. As a first observation, we all engage in the Lombard effect [2] by instinct when we increase our vocal effort to speak in a noisy environment. The increased intelligibility in our Lombard speech can be attributed to several changes including a louder and slower-paced speech, and a shift in spectral energy from lower to higher frequencies (spectral tilt flattening) [3]. Similar strategies can be adopted for the preprocessing of noise-free speech signals for subsequent listening in a noisy environment. However, changing the pace of the speech is not compatible with real-time processing. Furthermore, amplifying all parts of the audio signal equally is not efficient and can cause discomfort or harm to auditory systems. In this regard, volume adjustment is better left to listeners. To improve speech intelligibility beyond these preliminary considerations, a deeper understanding of what constitutes intelligibility would be necessary. Past research has found that vowels carry more information than consonants in fluent speech [4, 5]. Others have improved intelligibility by boosting the consonants against the vowels, i.e., increasing the consonant-vowel intensity ratio [6, 7, 8, 9]. These conclusions do not contradict each other since an increase in the energy of the consonants would also increase the overall energy of the audio relative to noise, which improves intelligibility. Yet other researchers would not seek to distinguish between consonants and vowels but instead boost the transients in speech or reallocate the spectral energy toward higher frequencies [10, 11, 12, 13]. Their research results have shown that speech intelligibility can be improved even when the overall audio energy remains unchanged. However, most speech enhancement algorithms to date tend to be computationally expensive and are somewhat difficult to implement for real-time signal processing.

This paper is an extension to an earlier conference paper [14], which introduced a wavelet-based algorithm to improve speech intelligibility along with some preliminary results. Wavelets [15, 16, 17], due to their highly localized nature in both time and frequency, are particularly suitable for representing and capturing the transient events in a non-stationary signal such as audio [18, 10, 11]. The wavelet-based speech enhancement algorithm reported in [14] was an adaptation from previous audio applications [19, 20]. It can be implemented in $O(n)$ time, where $n$ is the length of the audio signal. The proposed algorithm is real-time implementable and simpler than other previously reported algorithms. This paper further elaborates on the evaluation of speech intelligibility and the optimization scheme for the algorithm. The full data set will be presented along with analytical results that describe the algorithm's capabilities in three distinct scenarios: (1) enhancing noise-free speech, (2) enhancing speech corrupted by noise, and (3) enhancing speech for people with hearing loss.

The first results to be presented will be the preprocessing of noise-free speech for subsequent listening in a noisy environment, where intelligibility improvements under various background interference are shown. Next, for speech already corrupted by noise, improving intelligibility is a rather challenging task. It was reported in 2011 that speech enhancement algorithms could mainly improve speech quality but not intelligibility [21]. The main challenge lies in the difficulty in predicting nonstationary noise. Moreover, algorithms back then were not designed to improve speech intelligibility and as a result, noise could be reduced but the associated distortions degraded intelligibility. The present algorithm, on the contrary, specifically targets intelligibility improvement. The enhancement results on noisy speech will be presented.

The final demonstration is speech enhancement for listeners with hearing loss. There are two major types of hearing loss: conductive and sensorineural; both result in an attenuation of the incoming sound. However, the latter may further cause progressive hearing loss at higher frequencies, loudness recruitment, and poor temporal and frequency resolution, all of which can be highly detrimental to speech intelligibility [22]. For potential hearing aid applications, the present algorithm can partially compensate for the frequency-dependent hearing loss by shifting spectral energy to higher frequencies. Afterward, a dynamic range controller can provide the necessary amplification and loudness control for hearing aid fitting.

## 2. SPEECH ENHANCEMENT ALGORITHM

Using orthogonal wavelets as bases, a discrete-time audio signal can be broken down into and hence represented by its constituent wavelets. In practice, rather than dealing directly with the wavelet functions, the decomposition is implemented as a finite-impulse-response (FIR) filter bank operation. In the discrete wavelet transform (DWT), filter coefficients are defined for a pair of low- and high-pass decomposition filters associated with the wavelet. The audio signal is split into two halves by these two filters, followed by ↓2 downsampling. The filtered and decimated low- and high-pass signals are known as the approximation (cA) and detail (cD) coefficients, respectively. The entire process of splitting and decimation can be applied to cA, cD, and their offspring successively and selectively using the same pair of filters. In the most common pyramidal implementation of the Mallat algorithm [23], only the cA's are split further, thus generating a wavelet packet tree and an overlapping octave filter bank structure as shown in Fig. 1. Each additional level of decomposition offers a different perspective in time and scale/frequency. The net result is a multi-resolution wavelet representation of the audio signal. Reconstruction is the reverse of decomposition, and the inverse discrete wavelet transform (IDWT) is performed by using a pair of low- and high-pass reconstruction filters associated with the wavelet. Besides the convolution-based filter bank implementation, the wavelet decomposition and reconstruction can be efficiently implemented in real-time by lifting-based techniques [24, 25].

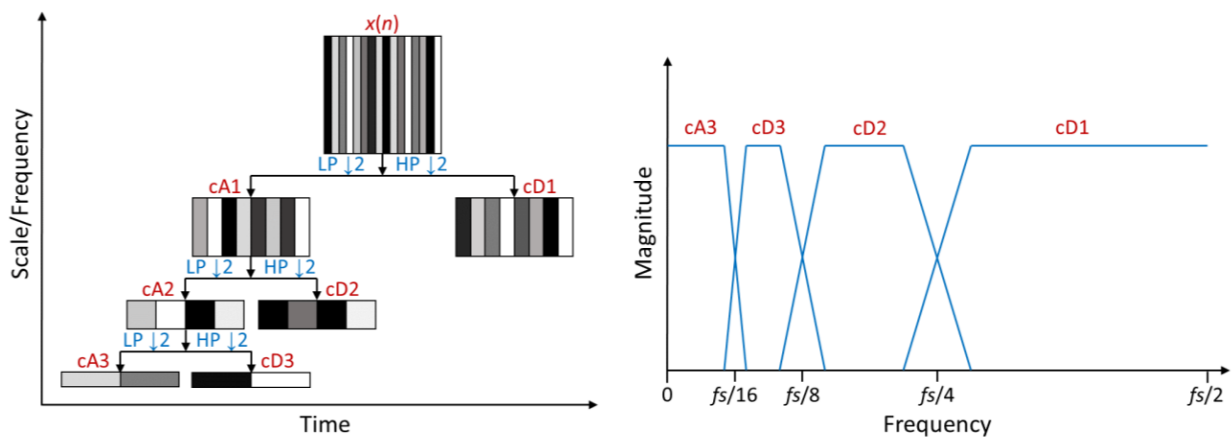

*Figure 1. (Left) Level-3 dyadic decomposition of a discrete-time audio signal x(n) into a wavelet packet tree via cascaded low-pass (LP) and high-pass (HP) finite-impulse-response (FIR) filtering followed by ↓2 downsampling, thereby achieving multi-resolution in time and scale/frequency. (Right) The resulting overlapping octave filter bank and their schematic magnitude spectra, fs is the sampling frequency.*

In the proposed algorithm shown in Fig. 2, a level-5 DWT is used to split the discrete-time speech signal into the following sub-bands: cA5, cD5, cD4, cD3, cD2 and cD1 in the order of increasing frequency. The sym12 wavelet [26] is chosen for its near symmetric properties, which provides a fairly compact representation of the audio waveform, and a relatively short filter of length 24, which makes the latency acceptable for real-time signal processing. At a sampling frequency of 44.1 kHz, it is found that cA5 mainly contains vowels and almost no consonants, and hence good separation is obtained. For other sampling frequencies, the level of the DWT can be adjusted accordingly. Various preset gains (or attenuations) are then applied to the sub-band signals before the signals are recombined via IDWT to form a modified version of the speech signal. This is followed by dynamic range compression, if necessary, which can limit the peak values of the modified signal to avoid clipping.

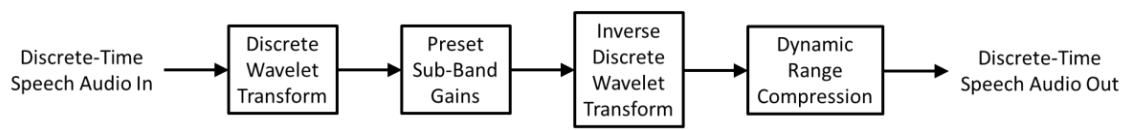

*Figure 2. Proposed real-time wavelet-based algorithm for improving speech intelligibility.*

## 3. SPEECH INTELLIGIBILITY EVALUATION

An objective and quantitative evaluation of speech intelligibility is crucial for the optimization of the sub-band gains. In the previous research of speech algorithms, evaluations were mainly performed on nonsensical syllables or individual words by human listeners. This, however, did not address the comprehensibility of speech made up of full sentences. Human listeners no doubt have the best say about intelligibility but suffer from objectivity and fatigue issues as the variants of the same speech are listened for many times. This might be overcome statistically by having enough human listeners who should ideally be native speakers for each language evaluated. However, doing so is impractical during the optimization and fine-tuning phase of the new algorithm.

To evaluate the intelligibility of full-sentence speech in multiple languages objectively and quantitatively, the Google Speech-to-Text recognition [27], which is based on machine learning and supports over 100 languages and variants, has been adopted in this research. The transcribed speech ("Google transcription") is compared with the correct text and the resulting percentage transcription accuracy is regarded as a quantitative measure of speech intelligibility. Since Google prefers lossless audio for best results and yet much of what is being transmitted in the media is compressed, there is a need to investigate how audio compression affects the Google transcription. As shown in Fig. 3, clean speech audios in 4 languages by both female and male speakers are compressed from the original lossless WAV to the lossy MP3 format at various bit rates. The results showed that the accuracy of the Google transcription is not affected once the bit rate exceeds 96 kbps for 2-channel audio. This condition is easily met by today's standards. With Google Speech-to-Text recognition, the transcription accuracy for noise-free speech is consistently above 90%. The addition of noise to speech can substantially degrade the transcription accuracy. Much like human listeners, when part of the speech is not clearly understood, Google tries to come up with the best guess based on commonly used terms, phrases, and topics. However, a distinctive advantage is the repeatability of the Google transcription accuracy, which stays within ±10 percentage points in most cases. In this regard, Google is arguably more reliable than human listeners are. Hence, Google speech-to-text recognition is considered a good choice for the evaluation of speech intelligibility in the present study.

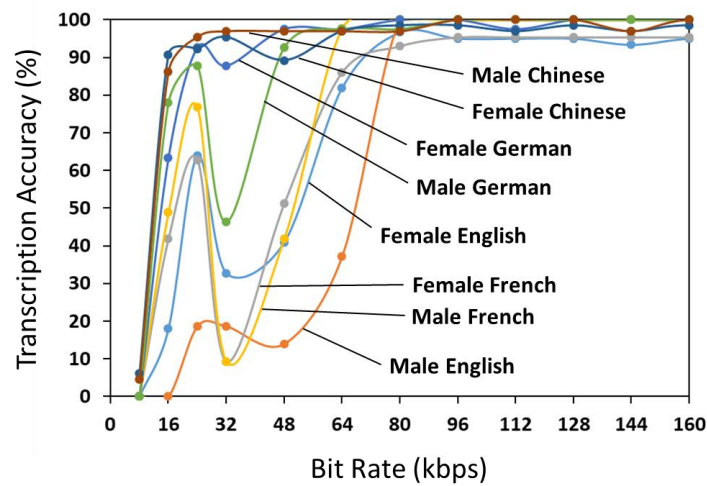

*Figure 3. Google Speech-to-Text transcription accuracy vs. the stereo bit rate of MP3 compression for English, French, Chinese and German speech examples by both female and male speakers. The original speech audios were in lossless WAV format before compression.*

## 4. OPTIMIZATION OF PRESET SUB-BAND GAINS

To come up with the optimized preset gains for the proposed speech enhancement algorithm in Fig. 2, a greedy search is performed on a set of sub-band gains that maximizes the Google transcription accuracy. The pseudocode for the optimization of the sub-band gains is shown in Fig. 4. The actual code is implemented in Python. The noise-free speech audio, which can be compressed or uncompressed, is input in the WAV format. Noise can be superimposed at various noise-to-signal ratios (NSR's). The use of the NSR on a linear scale rather than the more conventional signal-to-noise ratio (SNR) on a log scale in dB allows a clearer illustration of the speech degradation starting with the noise-free speech at NSR = 0.

Two different scenarios are considered. For the enhancement of noise-free speech for subsequent listening in a noisy environment, the speech audio is first enhanced at constant energy before mixing with noise and transcribing by Google. Then for the enhancement of noisy speech, the speech audio is first mixed with noise. Afterward, the noisy mix is enhanced at constant energy and transcribed by Google. Sequentially from cA5, cD5, cD4, cD3, cD2 to cD1, which is roughly in the order of decreasing audio energy, the gain for each sub-band is varied between 0 and 3 and that which generates the highest average transcription accuracy over the range of NSR from 0 to 3 (4.8 dB) is stored. At the start of the search, the sub-band gains are all set to one, which corresponds to the original speech. Subsequently, the gains can be set to that of the enhanced speech from the previous search. The iterative search is not exhaustive and can be terminated once definitive results are obtained. The best sub-band gain setting is used to construct the final enhanced speech, which is further checked by human listeners to make sure that it sounds natural and artifact-free.

```
Preset-Optimize (speech, noise):
  x ← speech waveform in WAV format
  n ← noise waveform in WAV format
  M ← array of preset gains, initialize to all 1's
  repeat
    for each band from cA5 to cD1:
      for gain from 0.0 to 3.0:
        M [band] ← gain
        for NSR from 0.0 to 3.0:
          if enhancing noise-free speech:
            mix ← Wavelet-Enhance (x, M) + NSR × n
          else if enhancing noisy speech:
            mix ← Wavelet-Enhance (x + NSR × n, M)
          score ← Google transcription of mix vs. ground truth
      choose M [band] = gain with highest overall average score
  until score does not improve
  return M

Wavelet-Enhance (speech, preset gains):
  x ← speech waveform in WAV format
  M ← array of preset gains
  bands from cA5 to cD1 ← DWT (x)
  bands-modified ← Multiply (bands, M)
  x-enhanced ← IDWT (bands-modified)
  x-enhanced-norm ← normalize x-enhanced to same energy as x
  if Peak (x-enhanced-norm) < threshold:
    return x-enhanced-norm
  else:
    return Compress (x-enhanced-norm)
```

*Figure 4. Pseudocode for the optimization of preset sub-band gains. Actual code is implemented in Python.*

# 5. NOISE-FREE SPEECH ENHANCEMENT RESULTS

In this study, noise-free speech is preprocessed for subsequent listening in a noisy environment. Two English and two Chinese speech examples, each 13 to 27 seconds long, are used to represent the two most spoken languages. Both female and male speakers are included for a better coverage of the evaluation. The contents of the speech are chosen to be ordinary topics in daily chats and free from specific terms or names that may otherwise bias the Google Speech-to-Text recognition. For a fair comparison of intelligibility, all 4 speech examples are normalized to the same rms value. As for the noisy environment, 4 types of interference are chosen: classical music, machine noise, babble, and Chinese pop song. These represent the range of background interference one typically encounters in daily life. For each of the 16 combinations, the normalized speech is superimposed with the corresponding interference at various NSR's from 0 to 3. The resulting audio mix is then transcribed by Google. The results for all 16 combinations of speech and interference are shown in Fig. 5a and 5b, in which the percentage transcription accuracy is plotted against the NSR. Labeled as "Unenhanced", this represents the baseline speech intelligibility under increasing background interference.

In the next step, for each specific NSR, the normalized noise-free speech is enhanced at constant energy before mixing with the background interference using a limited version of the sub-band gain optimization scheme outlined in Fig. 4. Since the sub-band gains are tailored for each specific NSR, the resulting enhancement is shown as "Point-to-Point" in Fig. 5a and 5b. This represents the best transcription accuracy that the sub-band gain search algorithm can achieve for a particular NSR. However, the level of the background interference often varies over time and cannot be determined in advance. There is therefore a need to enhance the speech with a universal set of sub-band gains that is optimized for the entire range of NSR of interest using the full optimization scheme in Fig. 4. The results of this enhancement, labeled as "Universal", are shown in Fig. 5a and 5b. Improvements in the transcription accuracy vary with the speaker, interference, and NSR in general. Marked improvements, as much as 86.7 percentage points, are evident in several speech and interference combinations at immediate NSR. Furthermore, for each combination of speech and interference, there appears to be a set of sub-band gains that can improve the speech intelligibility over a wide range of NSR, with the transcription accuracy close to the ceiling set by the point-to-point optimization at each specific NSR. This is encouraging since the optimized universal sub-band gains can then be preset for a particular speaker and situation, thus enabling the proposed wavelet-based algorithm for improving speech intelligibility to be implemented in real time.

Table 1 shows the optimized universal sub-band gains for all 16 speech and interference combinations. Overall, there is a reallocation of spectral energy toward the mid-frequency sub-bands, namely cD3 and cD4, for which the sub-band gains are invariably higher than one. This is consistent with the Lombard effect and prior research results. The results obtained here further suggest possible differences between speakers and languages. For example, the ratio between the two lowest frequency sub-band gains, i.e., cD5/cA5, in the enhanced Chinese speech is noticeable higher than that for the English speech.

Table 2 summarizes the noise-free speech enhancement results for all 16 speech and interference combinations using the optimized universal sub-band gains in Table 1. The overall improvement in the Google transcription accuracy is 16.9 percentage points on average. Among the 4 types of background interference, machine noise and babble appear to be the most detrimental, and classical music the least detrimental to speech intelligibility. Because the unenhanced transcription accuracy is high, the room for further enhancement is somewhat limited in the classical music example. Other than that, the biggest improvements in the transcription accuracy are seen with the babble and Chinese pop song interference and the male Chinese speaker.

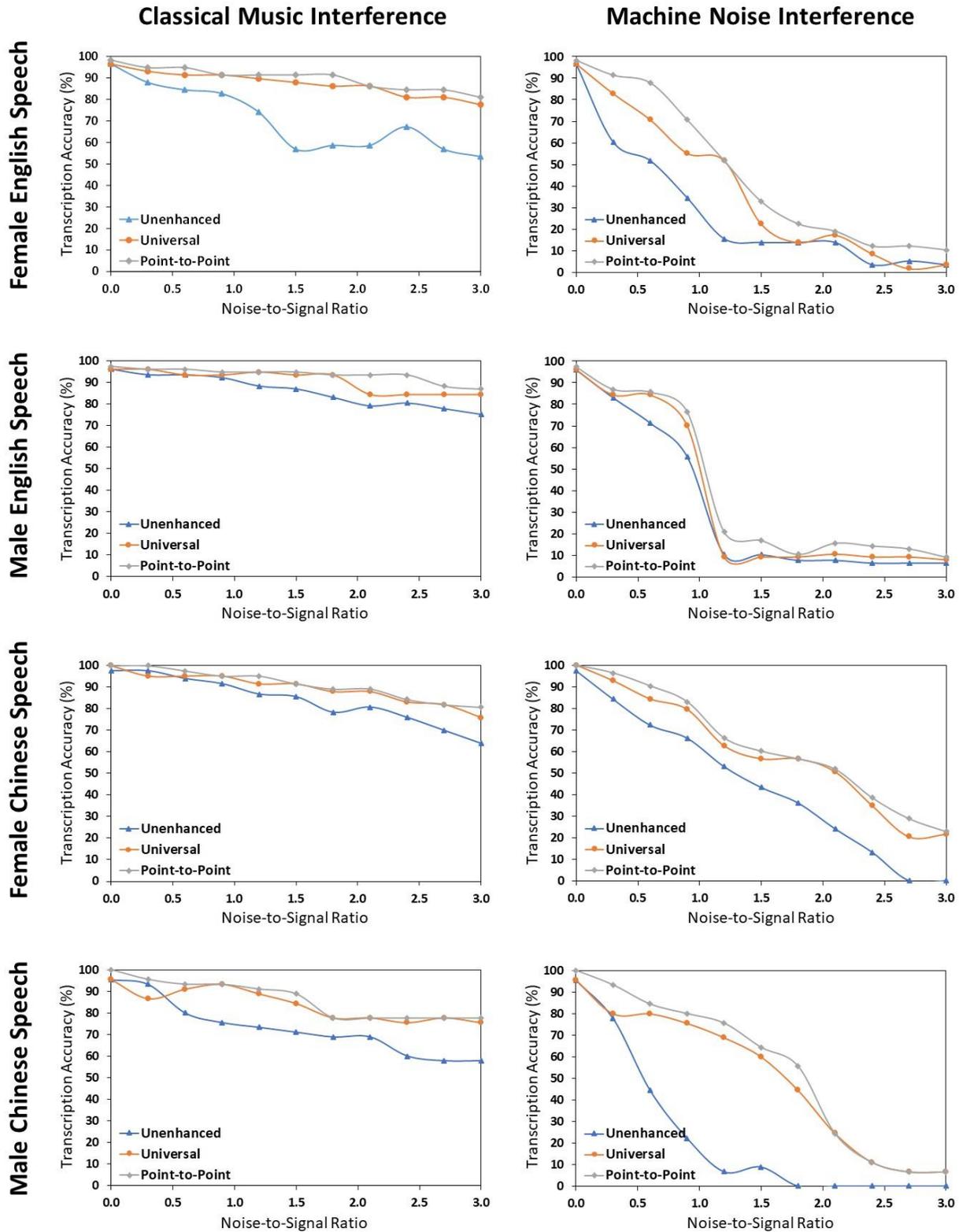

*Figure 5a. Enhancement of noise-free speech for subsequent listening in a noisy environment. Normalized noise-free speech is enhanced before mixing with background interference at various noise-to-signal ratios (NSR's). Afterward, the Google Speech-to-Text transcription accuracy of the mix containing the (i) unenhanced, (ii) point-to-point enhanced, and (iii) universally enhanced speech are plotted against the NSR for the first 8 combinations of speech (in rows) and interference (in columns).*

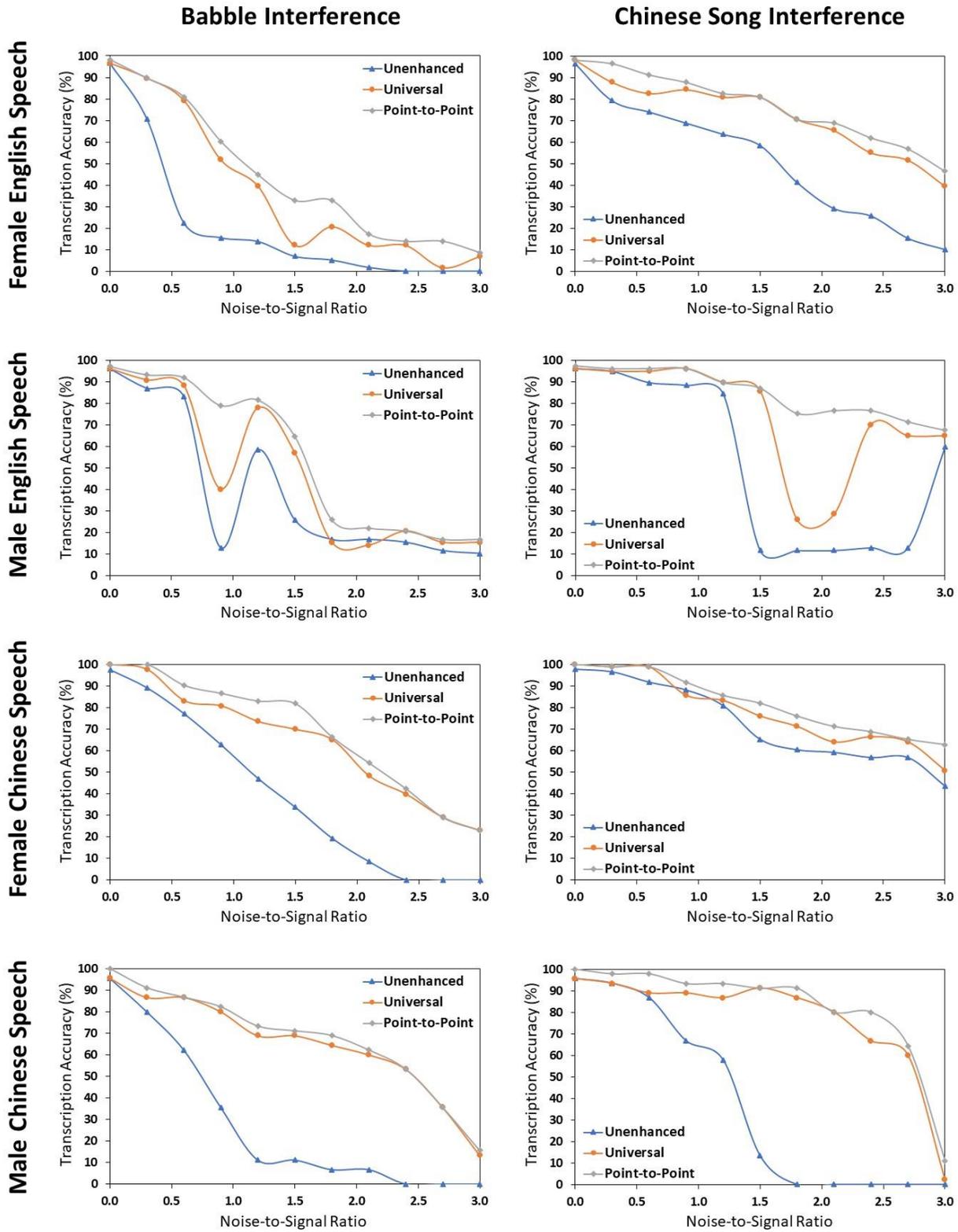

*Figure 5b. Enhancement of noise-free speech for subsequent listening in a noisy environment. Normalized noise-free speech is enhanced before mixing with background interference at various noise-to-signal ratios (NSR's). Afterward, the Google Speech-to-Text transcription accuracy of the mix containing the (i) unenhanced, (ii) point-to-point enhanced, and (iii) universally enhanced speech are plotted against the NSR for the second 8 combinations of speech (in rows) and interference (in columns).*

Table 1. Universal sub-band gains for the enhancement of noise-free speech for subsequent listening in a noisy environment. The 4 speech examples comprise female, male, English, and Chinese speakers. These are matched up against 4 types of background interference: classical music, machine noise, babble, and Chinese pop song. The sub-band gains are optimized for noise-to-signal ratios (NSR's) from 0 to 3.

|  | Classical Music | | | | | | Machine Noise | | | | | | Babble | | | | | | Chinese Song | | | | | |
| --- | --- | --- | --- | --- | --- | --- | --- | --- | --- | --- | --- | --- | --- | --- | --- | --- | --- | --- | --- | --- | --- | --- | --- | --- |
| Sub-Band | cA5 | cD5 | cD4 | cD3 | cD2 | cD1 | cA5 | cD5 | cD4 | cD3 | cD2 | cD1 | cA5 | cD5 | cD4 | cD3 | cD2 | cD1 | cA5 | cD5 | cD4 | cD3 | cD2 | cD1 |
| Female English | 1.0 | 0.5 | 2.1 | 3.1 | 0.3 | 0.5 | 1.1 | 0.4 | 1.8 | 3.1 | 1.5 | 3.3 | 1.1 | 0.3 | 1.7 | 4.2 | 0.0 | 0.6 | 1.0 | 0.6 | 2.5 | 2.5 | 0.2 | 2.1 |
| Male English | 1.1 | 0.4 | 1.2 | 1.5 | 0.1 | 0.0 | 1.0 | 0.7 | 1.2 | 1.7 | 0.0 | 0.0 | 0.9 | 0.9 | 1.5 | 1.5 | 0.0 | 0.0 | 1.0 | 0.5 | 1.2 | 2.2 | 0.2 | 0.0 |
| Female Chinese | 0.8 | 1.3 | 3.1 | 1.8 | 1.6 | 0.0 | 0.8 | 2.7 | 2.1 | 1.9 | 0.0 | 0.8 | 0.8 | 2.4 | 2.4 | 2.0 | 0.0 | 0.4 | 0.9 | 1.4 | 1.1 | 2.5 | 0.0 | 0.0 |
| Male Chinese | 0.7 | 3.0 | 3.7 | 3.7 | 3.7 | 3.7 | 0.7 | 2.7 | 4.7 | 6.0 | 0.0 | 2.0 | 0.8 | 2.3 | 3.8 | 5.3 | 4.5 | 0.0 | 0.7 | 2.1 | 4.2 | 4.9 | 8.5 | 3.5 |

Table 2. Noise-free speech enhancement results for all 16 speech and interference combinations using the optimized universal sub-band gains in Table 1. The Google Speech-to-Text transcription accuracy is averaged over the range of NSR from 0 to 3.

| Average Transcription Accuracy (%) | | Background Interference | | | | Speaker Average |
| --- | --- | --- | --- | --- | --- | --- |
| Speech | Clean Speech Enhancement | Classical Music | Machine Noise | Babble | Chinese Song | |
| Female English | Unenhanced | 70.7 | 28.4 | 21.2 | 51.3 | 42.9 |
|  | Enhanced | 87.5 | 38.6 | 38.4 | 72.6 | 59.2 |
|  | Improvements | 16.8 | 10.2 | 17.2 | 21.3 | 16.4 |
| Male English | Unenhanced | 86.1 | 32.9 | 39.6 | 52.2 | 52.7 |
|  | Enhanced | 90.8 | 36.3 | 48.4 | 73.8 | 62.3 |
|  | Improvements | 4.7 | 3.3 | 8.9 | 21.6 | 9.6 |
| Female Chinese | Unenhanced | 83.8 | 44.6 | 39.5 | 72.3 | 60.1 |
|  | Enhanced | 89.6 | 60.0 | 64.5 | 78.0 | 73.0 |
|  | Improvements | 5.8 | 15.4 | 25.0 | 5.7 | 13.0 |
| Male Chinese | Unenhanced | 72.9 | 23.2 | 28.1 | 37.6 | 40.5 |
|  | Enhanced | 84.0 | 50.3 | 64.9 | 76.4 | 68.9 |
|  | Improvements | 11.1 | 27.1 | 36.8 | 38.8 | 28.4 |
| Interference Average | Unenhanced | 78.4 | 32.3 | 32.1 | 53.3 | 49.0 |
|  | Enhanced | 88.0 | 46.3 | 54.0 | 75.2 | 65.9 |
|  | Improvements | 9.6 | 14.0 | 22.0 | 21.9 | 16.9 |

## 6. NOISY SPEECH ENHANCEMENT RESULTS

In this study, the enhancement of speech already corrupted by noise will be presented. The same 16 speech and interference combinations are used as before. The only difference is that, since the speech and interference are already mixed, the wavelet-based enhancement algorithm will process not just the clean speech, but the background interference as well. To proceed, a normalized noise-free speech is first mixed with a background interference at various NSR's from 0 to 3. The resulting audio mix is then transcribed by Google Speech-to-Text and the percentage accuracy vs. the NSR is shown as the "Unenhanced" plot in Fig. 6a and 6b. This represents the baseline speech intelligibility and is identical to that in Fig. 5a and 5b. In the next step, for each specific NSR, the noisy speech mix is enhanced at constant energy using a limited version of the sub-band gain optimization scheme outlined in Fig. 4. Since the sub-band gains are tailored for each specific NSR, the resulting enhancement is labeled as "Point-to-Point" in Fig. 6a and 6b. This represents the best transcription accuracy that the sub-band gain search algorithm can achieve for a particular NSR. However, the level of the background interference often varies over time and cannot be determined in advance. There is therefore a need to enhance the noisy speech with a universal set of sub-band gains that is optimized for the entire range of NSR of interest using the full optimization scheme in Fig. 4. The results of this enhancement, labeled as "Universal", are shown in Fig. 6a and 6b. Again, for each combination of speech and interference, there exists a universal set of sub-band gains that can improve the speech intelligibility over a wide range of NSR, with the transcription accuracy approaching the point-to-point optimization at each specific NSR.

In comparison with the previous enhancement of noise-free speech in the same examples, the transcription accuracy improvements for noisy speech are more remarkable with the male English speaker (as much as 71.4 percentage point improvement) and less remarkable with most of the other combinations.

Table 3 shows the optimized universal sub-band gains for all 16 speech and interference combinations. In contrast to the results for noise-free speech, only a few sub-bands are augmented. For the other sub-bands, the optimization algorithm does not find it helpful to boost the speech and noise together in the noisy speech mix. As a result, the achievable improvements in transcription accuracy are very much dependent on the extent of non-overlap between speech and interference energies in individual sub-bands. Table 4 summarizes the noisy speech enhancement results for all 16 speech and interference combinations using the optimized universal sub-band gains in Table 3. The overall improvement in the transcription accuracy is 9.5 percentage points on average versus 16.9 percentage points for the noise-free speech case. This is encouraging as the enhancement of noisy speech is a challenging task. In particular, the male English speaker, which shows the least improvement (9.6 percentage points on average) in the noise-free speech case, is now showing the most improvement (19.7 percentage points on average).

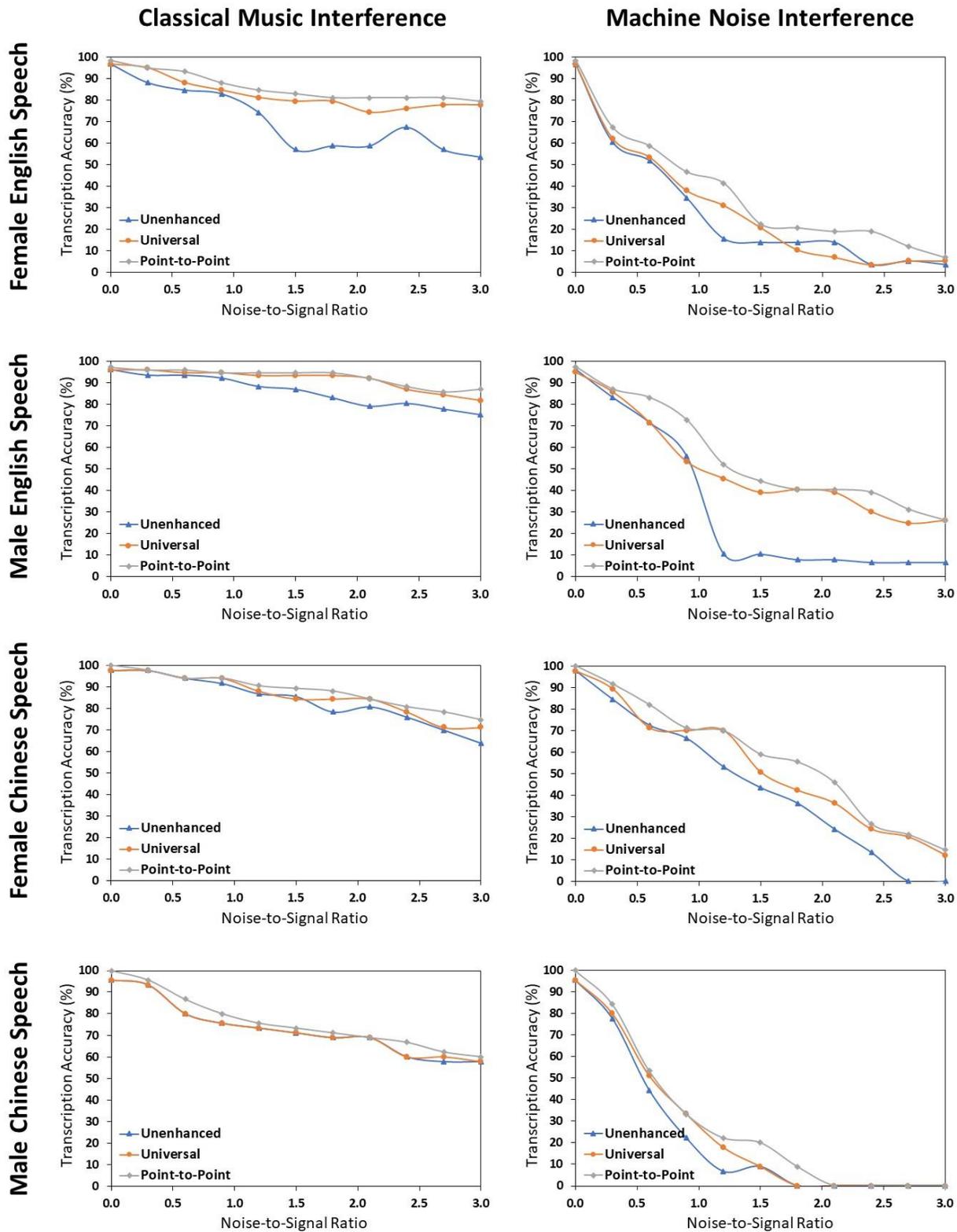

*Figure 6a. Enhancement of speech already corrupted by noise. Normalized noise-free speech is first mixed with background interference at various noise-to-signal ratios (NSR's). Afterward, the Google Speech-to-Text transcription accuracy of the (i) unenhanced, (ii) point-to-point enhanced, and (iii) universally enhanced mix are plotted against the NSR for the first 8 combinations of speech (in rows) and interference (in columns).*

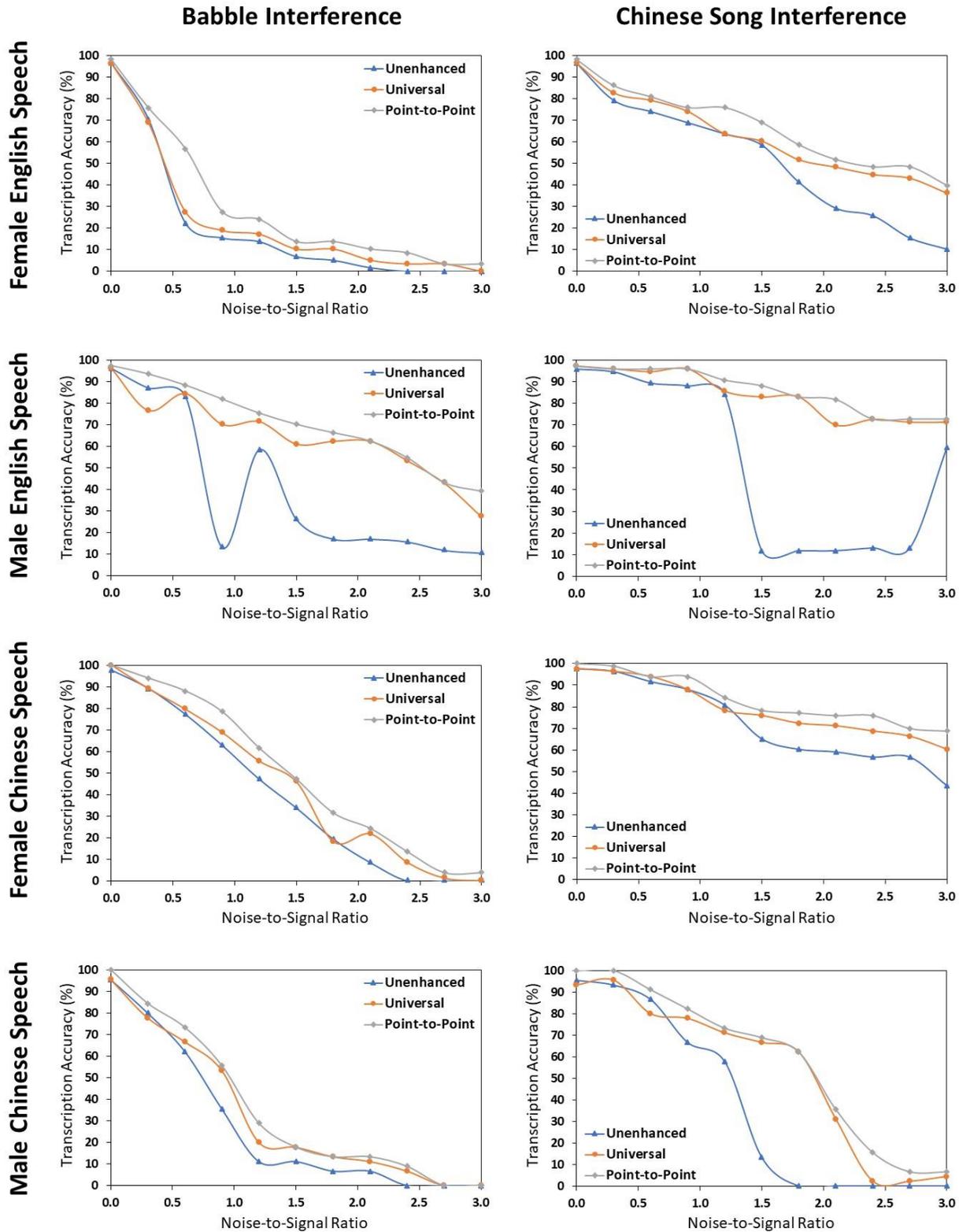

*Figure 6b. Enhancement of speech already corrupted by noise. Normalized noise-free speech is first mixed with background interference at various noise-to-signal ratios (NSR's). Afterward, the Google Speech-to-Text transcription accuracy of the (i) unenhanced, (ii) point-to-point enhanced, and (iii) universally enhanced mix are plotted against the NSR for the second 8 combinations of speech (in rows) and interference (in columns).*

*Table 3. Universal sub-band gains for the enhancement of speech already corrupted by noise. The 4 speech examples comprise female, male, English, and Chinese speakers. These are matched up against 4 types of background interference: classical music, machine noise, babble, and Chinese pop song. The sub-band gains are optimized for noise-to-signal ratios (NSR's) from 0 to 3.*

| | Classical Music | | | | | | Machine Noise | | | | | | Babble | | | | | | Chinese Song | | | | | |
|---|---|---|---|---|---|---|---|---|---|---|---|---|---|---|---|---|---|---|---|---|---|---|---|---|
| Sub-Band | cA5 | cD5 | cD4 | cD3 | cD2 | cD1 | cA5 | cD5 | cD4 | cD3 | cD2 | cD1 | cA5 | cD5 | cD4 | cD3 | cD2 | cD1 | cA5 | cD5 | cD4 | cD3 | cD2 | cD1 |
| Female English | 1.1 | 0.7 | 0.9 | 1.0 | 0.0 | 0.0 | 1.2 | 0.6 | 0.7 | 0.6 | 0.6 | 0.2 | 1.2 | 0.6 | 0.6 | 0.6 | 0.7 | 0.0 | 1.1 | 0.8 | 0.8 | 2.4 | 0.0 | 0.8 |
| Male English | 1.2 | 0.4 | 0.5 | 1.3 | 0.0 | 0.7 | 1.2 | 0.3 | 0.4 | 0.5 | 0.7 | 1.2 | 1.0 | 0.5 | 1.2 | 2.2 | 0.0 | 0.3 | 1.2 | 0.3 | 0.6 | 0.7 | 0.0 | 0.1 |
| Female Chinese | 1.0 | 0.8 | 0.8 | 1.0 | 1.6 | 2.4 | 1.1 | 0.5 | 0.2 | 0.4 | 0.0 | 0.4 | 0.9 | 0.9 | 1.6 | 2.7 | 0.7 | 1.6 | 1.0 | 0.9 | 1.1 | 0.7 | 0.0 | 1.7 |
| Male Chinese | 1.0 | 1.0 | 1.0 | 1.0 | 1.0 | 0.6 | 1.0 | 1.2 | 1.2 | 1.2 | 1.2 | 1.2 | 1.0 | 0.7 | 1.5 | 3.2 | 0.5 | 0.0 | 1.0 | 0.5 | 0.5 | 0.3 | 0.0 | 0.0 |

*Table 4. Noisy speech enhancement results for all 16 speech and background combinations using the optimized universal sub-band gains in Table 3. The Google Speech-to-Text transcription accuracy is averaged over the range of NSR from 0 to 3.*

| Average Transcription Accuracy (%) | | Background Interference | | | | Speaker Average |
|---|---|---|---|---|---|---|
| Speech | Noisy Speech Enhancement | Classical Music | Machine Noise | Babble | Chinese Song | |
| Female English | Unenhanced | 70.7 | 28.4 | 21.2 | 51.3 | 42.9 |
| | Enhanced | 82.6 | 30.3 | 23.8 | 61.9 | 49.6 |
| | Improvements | 11.9 | 1.9 | 2.7 | 10.7 | 6.8 |
| Male English | Unenhanced | 86.1 | 32.9 | 39.6 | 52.2 | 52.7 |
| | Enhanced | 91.6 | 49.9 | 64.3 | 83.8 | 72.4 |
| | Improvements | 5.6 | 17.0 | 24.8 | 31.7 | 19.7 |
| Female Chinese | Unenhanced | 83.8 | 44.6 | 39.5 | 72.3 | 60.1 |
| | Enhanced | 85.9 | 53.0 | 44.4 | 79.0 | 65.6 |
| | Improvements | 2.1 | 8.4 | 4.8 | 6.7 | 5.5 |
| Male Chinese | Unenhanced | 72.9 | 23.2 | 28.1 | 37.6 | 40.5 |
| | Enhanced | 73.1 | 26.1 | 32.9 | 53.3 | 46.4 |
| | Improvements | 0.2 | 2.8 | 4.9 | 15.8 | 5.9 |
| Interference Average | Unenhanced | 78.4 | 32.3 | 32.1 | 53.3 | 49.0 |
| | Enhanced | 83.3 | 39.8 | 41.4 | 69.5 | 58.5 |
| | Improvements | 4.9 | 7.5 | 9.3 | 16.2 | 9.5 |

.

# 7. SPEECH ENHANCEMENT FOR PEOPLE WITH HEARING LOSS

The wavelet-based algorithm has been explored for its potential to improve speech intelligibility for people with hearing loss. Hearing loss is typically assessed by pure-tone audiometry in which the increase in hearing threshold levels compared to people of normal hearing, expressible in dB hearing loss, is plotted against frequency on an audiogram [29]. To automate the speech intelligibility optimization, a hearing loss simulator has been developed that takes into account: (i) the absolute threshold of hearing for normal hearing, (ii) the dB hearing loss from an audiogram, and (iii) the loudness recruitment, which is a condition associated with sensorineural hearing loss that is characterized by an abnormal and rapid growth in loudness as sound intensity increases [22]. A discrete-time noise-free speech file is first processed by the hearing loss simulator, after which it becomes highly attenuated especially at high frequencies. The processed speech signal is then renormalized to the original signal energy before being transcribed by Google Speech-to-Text for a fair comparison of the transcription accuracy. As expected, the transcription accuracy can drop significantly below 100% after the hearing loss simulation. This is a direct result of missing high frequency consonants, which makes the speech audio less intelligible, and which cannot be mitigated by a simple amplification of the entire audio. Next, the original speech signal is enhanced at constant energy using the sub-band gain search algorithm previously described prior to the hearing loss simulation. The results for a simulated case of moderate sensorineural hearing loss are presented in Fig. 7. A new female English speech example is chosen. The original speech audio has a Google transcription accuracy of 100%. After the hearing loss simulator, the accuracy drops to 9.5%. The optimally enhanced speech has a higher concentration of energy in the mid-frequency sub-bands. The pre-simulator transcription accuracy is still 100% whereas the post-simulator accuracy is improved to 64%.

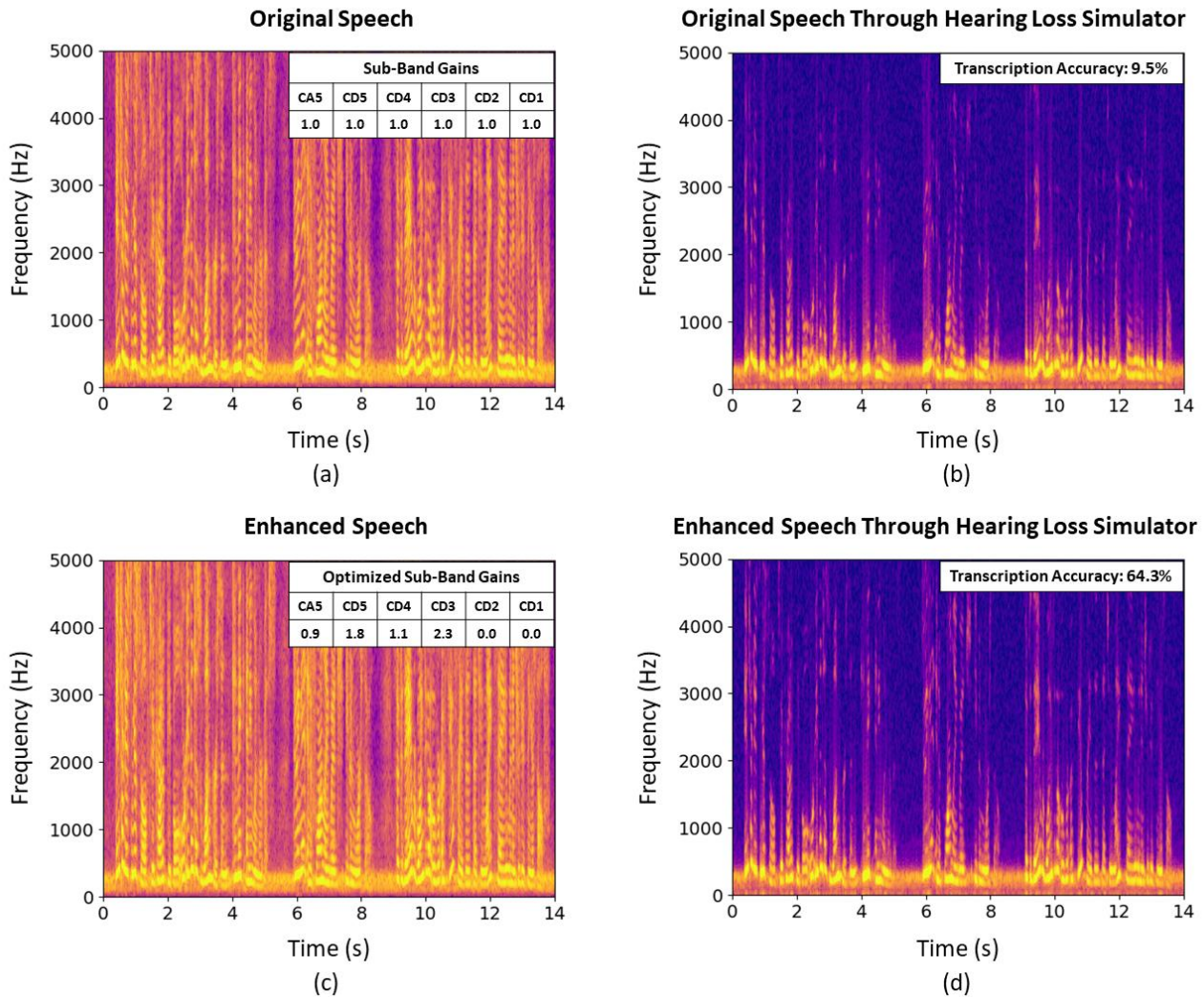

*Figure 7. Intelligibility improvement for listeners with hearing loss. Normalized spectrograms of (a) the original female English speech, (b) original speech processed by the hearing loss simulator, (c) enhanced speech, and (d) enhanced speech processed by the hearing loss simulator. The original speech is enhanced at constant energy. The enhanced speech has a higher concentration of energy (appears brighter) in the mid-frequency sub-bands and demonstrates a post-simulator speech-to-text transcription accuracy of 64% vs. 9.5% for the original unenhanced speech. The simulated hearing loss is 30, 30, 40, 50, 60, 60, 60 dB at 125, 250, 500, 1000, 2000, 4000, 8000 Hz, respectively.*

## 8. DISCUSSION AND CONCLUSIONS

The optimization of a wavelet-based algorithm to improve speech intelligibility is described. The discrete-time speech signal is split into frequency sub-bands via a multi-level discrete wavelet transform. Various gains (or attenuations) are applied to the sub-band signals before they are recombined to form a modified version of the speech. The sub-band gains are adjusted while keeping the overall signal energy unchanged, and the speech intelligibility under various background interference and simulated hearing loss conditions is enhanced by a sub-band gain optimization scheme.

Speech intelligibility is evaluated using Google Speech-to-Text transcription, which is based on state-of-the-art deep learning neural network. There are several advantages in using speech recognition tools like Google Speech-to-Text. These tools enable the automated transcription of full-sentence speech in over 100 languages, thus avoiding the logistics of recruiting qualified human subjects for the arduous listening tests involved during the optimization and fine-tuning phase of the new algorithm. The objectivity and repeatability of machine speech recognition surpass human listeners. The resulting percentage transcription accuracy is of sufficient quality to allow the degradation pattern of speech intelligibility for different interference sources to be visually compared and analyzed.

The enhancement results of clean and noisy speech with the proposed wavelet-based algorithm are then presented. The two most spoken languages (English and Chinese) by both male and female speakers are chosen along with 4 common types of background interference (classical music, machine noise, babble, and pop song). The results show that the type of background interference matters, and that classical music is the least detrimental and babble and machine noise are the most detrimental to the intelligibility of the original unenhanced speech in the examples given.

For the enhancement of noise-free speech for subsequent listening in a noisy environment, the overall intelligibility is improved, and the Google transcription accuracy is increased by an average of 16.9 and as much as 86.7 percentage points by reallocating the spectral energy toward the mid-frequency sub-bands, effectively increasing the consonant-vowel intensity ratio. This is reasonable since the consonants are relatively weak and of short duration, which are therefore the most likely to become indistinguishable in the presence of background interference. The results are consistent with the Lombard effect and prior research. The enhanced speech sounds brighter in general but otherwise remains natural, artifact free, and agreeable to human listeners.

For speech already corrupted by background noise, improving intelligibility is challenging but still realizable with an increased transcription accuracy of 9.5 percentage points on average and 71.4 maximum. In this case, the wavelet algorithm will process not only the speech, but the background noise that can be dominating as well. As a result, the spectral energy reallocation pattern and achievable improvements are dependent on the extent of non-overlap between the speech and noise spectra. The enhanced noisy speech often sounds not much different from the original but the background interference is subdued. Further refinement is possible for an adaptive algorithm that can sample the background noise during the pauses in the speech to improve the real-time enhancement.

For each combination of speech and background interference, there appears to be a universal set of sub-band gains that can improve the speech intelligibility over a wide range of noise-to-signal ratio (NSR), with transcription accuracies close to the ceiling set by the point-to-point optimization at each specific NSR. This is encouraging since the level of background interference often varies over time and cannot be determined in advance. More importantly, the universal sub-band gains can be preset for a particular speaker and situation, thus enabling the wavelet-based algorithm to be implemented for real-time speech processing.

The proposed wavelet algorithm has also been explored for hearing loss applications. A hearing loss simulator has been developed in this study to model sensorineural hearing loss. Still subjected to the constraint of constant-energy speech enhancement, the wavelet-based algorithm has demonstrated the potential to improve speech intelligibility under simulated hearing loss conditions. More work will be conducted in the area of hearing loss compensation once the constant-energy constraint is removed.

The proposed wavelet-based algorithm for improving speech intelligibility is implementable for real-time signal processing and comparatively simpler than others reported. Potential applications include speech enhancement, hearing aids, machine listening, and a better understanding of speech intelligibility.

## ACKNOWLEDGMENT

The authors would like to thank the Undergraduate Research Opportunities Program (UROP) of The Hong Kong University of Science and Technology for supporting this work [grants UROP20EG01 and UROP21EG11].